\documentclass{aa}
\usepackage{graphicx}
\usepackage[varg]{txfonts}
\usepackage{natbib}
\usepackage{xcolor}
\usepackage[colorlinks=true, allcolors = blue]{hyperref}
\newlength\staretab

\begin{document}

\title{Horizontal flows in the atmospheres of chemically peculiar stars}

    \author{A. ud-Doula
          \inst{1}\fnmsep\thanks{corresponding author}
          \and
          J. Krti{\v{c}}ka
           \inst{2}
          \and
          B. Kub\'atov\'a
          \inst{3}
          }
   \institute{Penn State Scranton, 120 Ridge View Drive, Dunmore, PA 18512, USA\\
              \email{asif@psu.edu}
              \and
           Department of Theoretical Physics and Astrophysics,
           Masaryk University, Kotl\'a\v rsk\' a 2, CZ-611\,37 Brno, Czech
           Republic
           \and
           Astronomical Institute of the Czech Academy of Sciences,
Fri\v{c}ova 298, 251 65 Ond{\v{r}}ejov,
Czech Republic 
}

\date{Received}

%\titlerunning{Horizontal Breeze}
 %  \authorrunning{A. ud-Doula et al.}

\abstract
% Context
    {Classical chemically peculiar stars exhibit atmospheres that are often structured by the effects of atomic diffusion. As a result of these elemental diffusion and horizontal abundance variations, photospheric temperature varies at a given height in the atmosphere. This may lead to horizontal flows in the photosphere. In addition, the  suppression of such flows by magnetic field can alter the elemental transport processes.}
  % aims heading (mandatory)
   {Using a simplified model of such a structured atmosphere and 2D MHD simulations  of a typical He-rich star, we examine atmospheric flows in these chemically peculiar stars which often are strongly magnetic.} 
  % methods heading (mandatory)
   {We use \texttt{Zeus-MP} which is a Fortran 90 based publicly available parallel finite element modular code.}
  % results heading (mandatory)
   {We find that for non-magnetic stars of spectral type BA, atmospheric flow related to horizontal temperature gradient can reach $1.0\,\text{km}\,\text{s}^{-1}$ yielding mixing timescale of order of tens of days. For the magnetic counterparts, the flow speeds are an order of magnitude lower allowing for stratification of chemical elements.}
  % conclusions heading 
   {Magnetic field can influence the dynamics in atmospheres significantly. Strong horizontal magnetic field inhibits flow in the vertical direction, while strong vertical magnetic field can suppress horizontal atmospheric flow preventing elemental mixing.}


\maketitle

\section{Introduction}
\label{intro}

The atmospheres of classical chemically peculiar stars of the upper part of the
main sequence are structured by the effects of atomic diffusion under the
influence of the radiative and gravitational force \citep{mirivi,ads,alestimag}.
In absence of strong mixing, this leads to horizontal and vertical abundance
stratification in the outer layers of these stars
\citep{kowad,koc37776,rykochba}.

The outer layers of chemically peculiar stars are frequently stabilized by strong
magnetic field \citep{morbob,wamimes,grunmimes}. The rotation of stellar surface
strewn with in-homogeneously distributed elements together with frozen-in
magnetic field manifests itself in plethora of remarkable phenomena. This includes
regular light variability \citep{falkal,labarcptess}, which stems from the
flux frequency redistribution induced by horizontal chemical inhomogeneity
\citep{mythetaaur,prvalis}, or radio activity originating from the
interaction of a weak stellar wind with strong magnetic field
\citep{myskal,osma,recrad}.

Although many detailed characteristics of chemically peculiar stars can be
understood from theoretical models, the processes shaping their surface
elemental distribution are not fully understood. The diffusion models predict a
tight correlation between surface elemental distribution and magnetic field
\citep{alestimag}, what is perhaps the case only for some elements
\citep{putting}.  Neither the observed characteristics of light variations of magnetic chemically peculiar stars do not seem to support a strong correlation between magnetic field and surface abundance distribution \citep{jagnemag}. 
In particular, simulations of light curves based on abundance spots associated with magnetic fields predict a much higher prevalence of double-wave light curves per rotation period than is observed.

%In particular, simulations of light curves from abundance spots connected with magnetic field predict much higher occurrence of light curves showing double wave per period than the observed fraction of these light curves.

Although the theoretical diffusion models and abundance distributions inferred
from observations seem to contradict, any additional process that modifies the
surface abundance distribution is likely to be rather weak. This is indicated by
stability of the observational characteristics of magnetic chemically peculiar
stars. Neither the observed light curves \citep{humkep} nor the underlying
surface abundance distribution \citep{popr} seem to show any secular changes.
This implies that either the abundance stratification has reached an equilibrium
state or that the forming processes operate on a longer timescale
\citep[e.g.,][]{alesido}.

A better understanding of the diffusion processes can be perhaps gleaned from the
analysis of chemically peculiar stars of Am and HgMn type. These stars do not
show strong magnetic fields \citep{makagmagne,tangne,catmagne,blazalmag} that
would impede the interaction of mixing processes and elemental diffusion unlike in the magnetic chemically peculiar stars.
This is likely the reason why the rotational light variability in HgMn stars is much weaker than in magnetic chemically peculiar stars. As a result, the light variability of a fraction of HgMn stars remained undetected until the age of precise space borne instruments \citep{kohgmntess}. Moreover, the connection between the light variability and surface abundance inhomogeneities is not well established. Despite this, in some of these stars it was possible to detect secular changes of surface elemental
distribution of yet unknown origin \citep{briquet2010,korphiphe,prvphiphe}.

While the deviations from the axial symmetry of the stellar surface either due
to the magnetic fields \citep{fulmat} or due to hydrostatic scale-height
variations caused by horizontal abundance variations \citep{myacen} are rather
small, other dynamical processes were invoked to provide additional mixing of
atmospheres of chemically peculiar stars. Such processes may include turbulence
and mass loss \citep{mirivi,alesvit} or meridional circulation \citep{Mich83}.

In addition to these processes, there is a possible mixing process directly
linked to peculiarity, which was not studied in detail up to now. As a result of
elemental diffusion and induced horizontal abundance variations, the
photospheric temperature varies at a given height in the atmosphere. Such
horizontal temperature variations exist regardless of the overall constancy of the
effective temperature \citep{molnar} and appear as a result of influence of
heavy element abundance on opacity. These temperature variations occurring in the
optically thin photospheric layers are rather weak (on the order of $10-100~\mathrm{K}$), and are connected with surface brightness variations and light
variability \citep{preslo,myhd37776,mycuvir}.

The variation in temperature at a given height within the atmospheres of chemically peculiar stars implies that pressure is likewise non-uniform at the same altitude \citep[e.g.][]{peter}. In such cases, horizontal pressure differences generate weak lateral flows analogous to atmospheric winds on Earth. Given the complex surface abundance patterns in chemically peculiar stars, the resulting flow induced by these pressure gradients represents a complex three-dimensional problem. Numerical modeling of this phenomenon is challenging due to the substantial difference in the horizontal and vertical extents of the flow, with significant disparities in scale making such simulations computationally intensive. Furthermore, the structure of the flow varies from star to star due to unique abundance stratifications across each stellar surface.

To develop a clearer physical understanding of this phenomenon, we address a simplified model assuming an initial linear temperature gradient in two dimensions directly over the region of chemical peculiarity. This approach enables us to estimate the intensity of the pressure gradient-driven flow while capturing essential dynamics. Additionally, we introduce uniform magnetic fields of different orientations and strengths into our simulations to explore the potential damping effects on these flows, which is particularly relevant in the atmospheres of the chemically peculiar magnetic stars, where magnetic suppression of flow could alter the transport processes.

In section \ref{methods} we discuss the specifics of our numerical methods. We then present our results in section \ref{simulation_results}. We follow this by a brief discussion in section \ref{Discussion} before a short summary and future work in section \ref{summary}.

\section{Numerical Methods}
\label{methods}

To model flows stemming from the horizontal temperature gradients, we set up a `box-in-a-star' simulation of the hydrostatic stellar atmosphere wherein a chemical spot with varying temperature is mimicked by a region with temperature $\Delta T$ higher than the effective temperature $T_\text{eff}$ of the star. 
%To maximize the effects of possible wind flows, we set  $\Delta~T=100~\mathrm{K}$.

We use  the publicly available hydrodynamic code \texttt{Zeus-MP} \citep{Hayes2006}  to model the atmospheric structure of a chemically peculiar star under a
planar approximation. The stellar parameters used in the simulations here are
chosen to approximate a typical magnetic helium-strong star HD~37776 \citep{brzda}  and are as follows: mass $M=8~M_{\odot}$, radius $R=4~R_{\odot}$, and effective temperature $T_{\rm eff}=22,000~\mathrm{K}$. Although we use full energy equation for the gas, we assume adiabatic index $\gamma=1.01$, reflecting nearly isothermal atmospheric conditions.

\begin{table}[h]
\caption{Summary of models with varying magnetic field strengths and orientations.}
\centering
\small
\begin{tabular}{c| c c c}
\hline\hline
\textbf{Model} & \textbf{Magnetic Field (G)} & \textbf{Field Orientation} \\
\hline
Non-magnetic & 0 & - \\
Weak Field & 1 & Horizontal, Vertical \\
Moderate Field & 10 & Horizontal, Vertical \\
Strong Field & 100 & Horizontal, Vertical \\
Very Strong Field & 1000 & Horizontal, Vertical \\
\hline
\end{tabular}
\label{Table:models}
\end{table}

\subsection{Initial and boundary conditions}
\label{Initial_conditions}

We initialize all our simulations with hydrostatic planar atmosphere with fixed gravitational acceleration $g~\equiv~G~M_\ast/R_\ast^2$ and base density $\rho_\ast=10^{-8}~\mathrm{g~cm}^{-3}$. At time $t=0~\mathrm{s}$ we introduce uniform magnetic field with different strengths and orientations as noted in Table \ref{Table:models}.

We include a  temperature variation in the form of a chemical spot, specified as
a linear temperature increase of $\Delta~T=100~\mathrm{K}$ at $X=0$, symmetric about
$X=0$ and spanning $5~H_\mathrm{scale}$ on both sides, where
$H_\mathrm{scale}= k_B T/(\mu m_H g)=a^2/g$ is the pressure scale height with isothermal
sound speed {  $a=\sqrt{k_B T/(\mu m_H)}~\sim~17\,\mathrm{km}\,\mathrm{s}^{-1}$ where $\mu$ is the mean molecular weight and $m_H$ is the hydrogen mass. For the parameters assumed in this work, it amounts to $H_\mathrm{scale}\sim 2.1 \times 10^8 \mathrm{cm}$.} %and orbital speed $v_\mathrm{orb} \sim 615\,\mathrm{km}\,\mathrm{s}^{-1}$.

For the horizontal ($X$) direction, we fix density and pressure to the value appropriate for a hydrostatic atmosphere. We also keep magnetic field components constant. While horizontal component of the velocity is kept at zero value in the interest of numerical stability, the vertical component is allowed to float through linear extrapolation from the two closest zones inside the computational domain. 

At the outer boundary, which represents the upper atmosphere, outflow boundary conditions are implemented, allowing all magnetohydrodynamic quantities to vary freely. This boundary is located at approximately $X=10~H_\mathrm{scale}$  or about 21,000 km above the stellar surface. 

Reflective boundary conditions are imposed along the vertical ($z$) direction, corresponding to the left and right sides in all subsequent figures. Under these conditions, the parallel components of velocity and magnetic field remain unchanged, while the perpendicular components reverse their direction.

The computational domain spans a total horizontal extent of \( 20~H_\mathrm{scale} \), while the vertical extent is half as large, measuring \( 10~H_\mathrm{scale} \).

The computational domain is discretized into \(128 \times 64\) grid zones along the \(x\)- and \(z\)-directions, yielding approximately 6--7 grid points per scale height (\(H_\mathrm{scale}\)). A total of nine models are explored: one non-magnetic reference case and eight magnetic configurations with field strengths of 1, 10, 100, and \(1000~\mathrm{G}\) in both horizontal and vertical orientations. Each simulation spans a physical time of \(100~\mathrm{ks}\), corresponding to over 50 sound crossing times, ensuring adequate time for the system to reach equilibrium.

\section{Simulation results} 
\label{simulation_results}

\subsection{Non-magnetic Model}
\label{sec:Non-magnetic}
\begin{figure*}
\includegraphics[width=\textwidth]{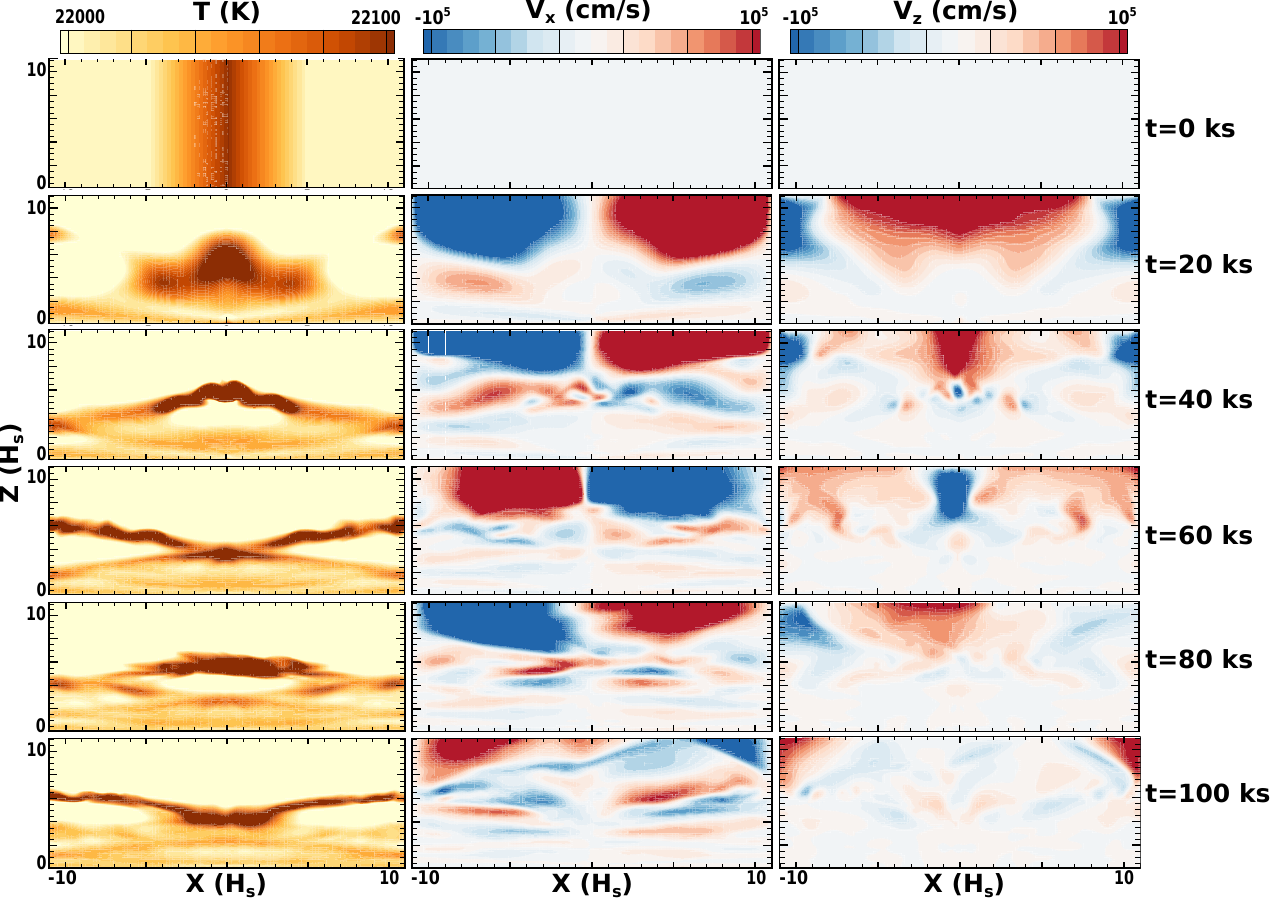}
\caption{Time evolution of the non-magnetic model. The top row shows the initial snapshot, with subsequent time snapshots taken every $20~\mathrm{ks}$. The left column presents temperature in Kelvin, the middle column illustrates horizontal flow velocity, and the right column displays vertical flow velocity in $\mathrm{cm}~\mathrm{s}^{-1}$.}
\label{fig-0G}
\end{figure*}

To establish a basis for comparison across all magnetic models, we first examine the non-magnetic case ($0~\mathrm{G}$). Fig.~\ref{fig-0G} illustrates the temporal evolution of the non-magnetic model, with panels showing temperature $T$ (left column), horizontal velocity $v_{x}$ (middle column), and vertical or radial velocity $v_{z}$ (right column), all in \texttt{cgs} units. Initially, the model is assumed to be in hydrostatic equilibrium along the vertical direction. However, the imposed temperature gradient across the modeled chemical spot quickly generates a horizontal pressure gradient, which initiates lateral flow that facilitates the mixing of cooler and warmer gas regions.

The fixed temperature gradient at the base, representing the stellar surface, likely sustains the circulation flow. This is
visually evident in the horizontal flow velocity (middle panels of Fig.~\ref{fig-0G}), where {red colour represents lateral motion of the gas to the right and blue represents motion to the left. Likewise, the red colour in the right panel represents upward flow while blue represents downward motion. Thus,} alternating red and blue regions indicate continuous flow
cycles. The temperature evolution of the left column highlights this circulation  further, as
pockets of hot gas ascend and then sink back, driven by the established pressure
and temperature gradients. The time-scale of the velocity alternation is given by
the time during which the flow moves from the centre of the computational domain
to its boundaries and bounces back, which is about $30~\mathrm{ks}$.

Throughout the simulation, typical flow velocities reach magnitudes on the order
of $10^5~\mathrm{cm}~\mathrm{s}^{-1}$, reflecting the strength of the pressure
gradient-driven flow in this non-magnetic setup. The velocities can be estimated
from the equation of motion for the subsonic flow $\rho\partial v_x/\partial
t=-\partial p/\partial x$, which on the adopted scale of temperature
inhomogeneities proportional to the length of the simulation box gives
velocities of about $v_x\sim a\sqrt{\Delta
T/T_\mathrm{eff}}\sim10^5~\mathrm{cm}~\mathrm{s}^{-1}$.

We have not shown the density structures of any of the models as they are essentially typical of hydrostatic atmosphere with exponential distribution which is not affected significantly by small velocity flows here.

\subsection{Standard $10~\mathrm{G}$ Horizontal Magnetic Model}
\label{sec:magnetic}

\begin{figure*}
\includegraphics[width=\textwidth]{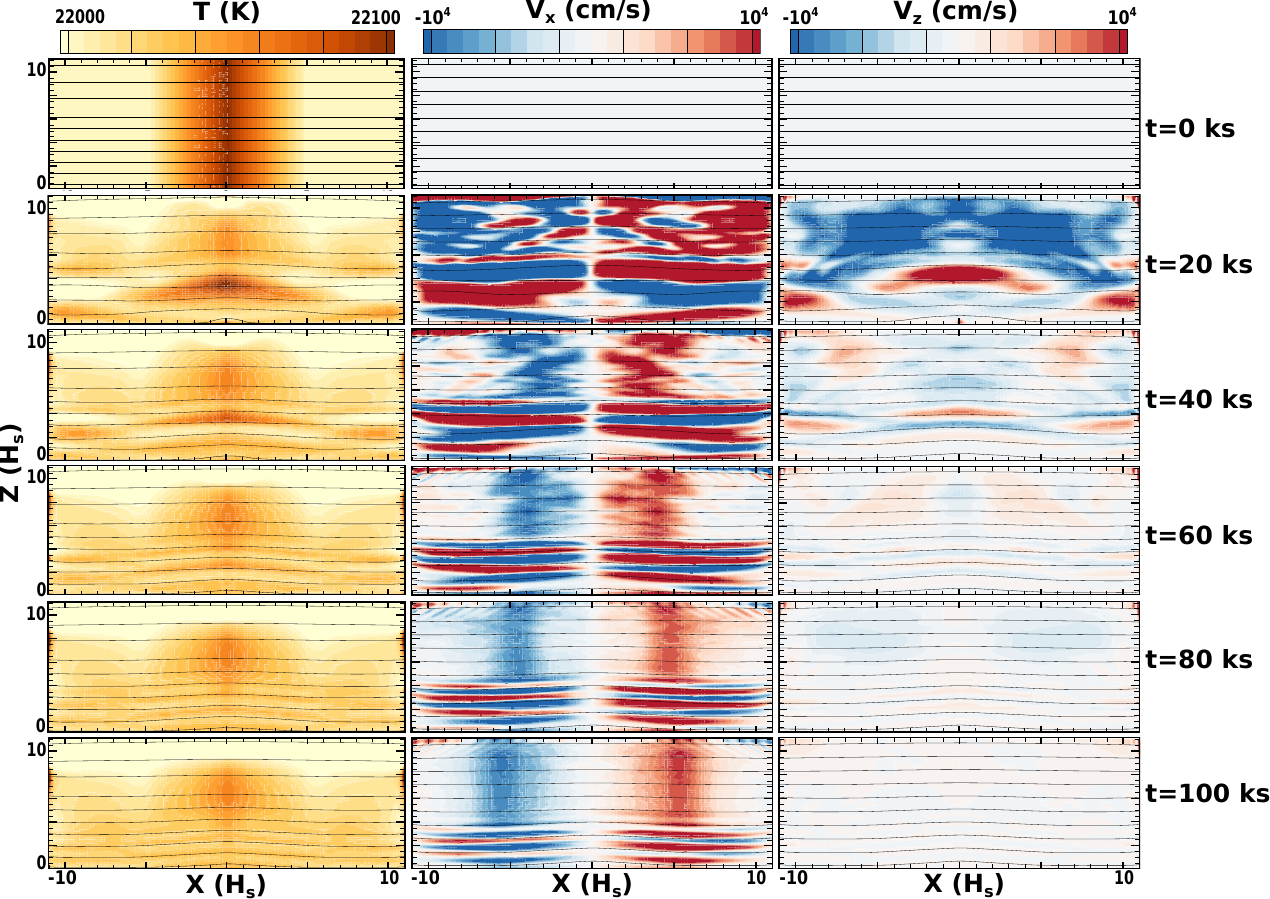}
\caption{Time evolution of the standard magnetic model with initial uniform $10~\mathrm{G}$ magnetic field (black lines) in the horizontal direction. The top row shows the initial time snapshot, with subsequent time snapshots taken every 20 ks. The left column shows temperature in Kelvin, the middle column illustrates horizontal flow velocity, and the right column displays vertical flow velocity in $\mathrm{cm}\,\mathrm{s}^{-1}$. Note that the range for velocity is factor 10 lower than for the non-magnetic case in Fig.~\ref{fig-0G}. This highlights  the effects magnetic field can have on the atmospheric flow.}
\label{fig-10G}
\end{figure*}

Fig. \ref{fig-10G} illustrates the time evolution of our standard magnetic model, which includes a $10~\mathrm{G}$ uniform horizontal magnetic field. In this model, the horizontal magnetic field is sufficiently strong to suppress vertical flows, significantly reducing the vertical mixing of gas. However, the magnetic field strength is not high enough to impede horizontal flows, allowing hot gas to spread laterally with speeds around $10^4~\mathrm{cm}~\mathrm{s}^{-1}$. Consequently, while a vertical temperature gradient is maintained, the hot gas is able to diffuse horizontally away from the initial chemical spot, leading to a more uniform temperature distribution across the stellar surface.

This behavior contrasts with cases involving stronger magnetic fields (see Table \ref{Table:models}), where both vertical and horizontal flows are heavily suppressed. In such high-magnetic field cases, the temperature stratification remains tightly confined to the chemical spot region due to the strong magnetic inhibition of flow. By comparing these cases, we gain insight into how varying magnetic field strengths influence the transport of energy and matter within the atmosphere, with intermediate fields allowing lateral diffusion while stronger fields create isolated, thermally stratified regions.

\subsection {Parameter Study}
\label{sec:parameter}
\begin{figure*}
\includegraphics[width=\textwidth]{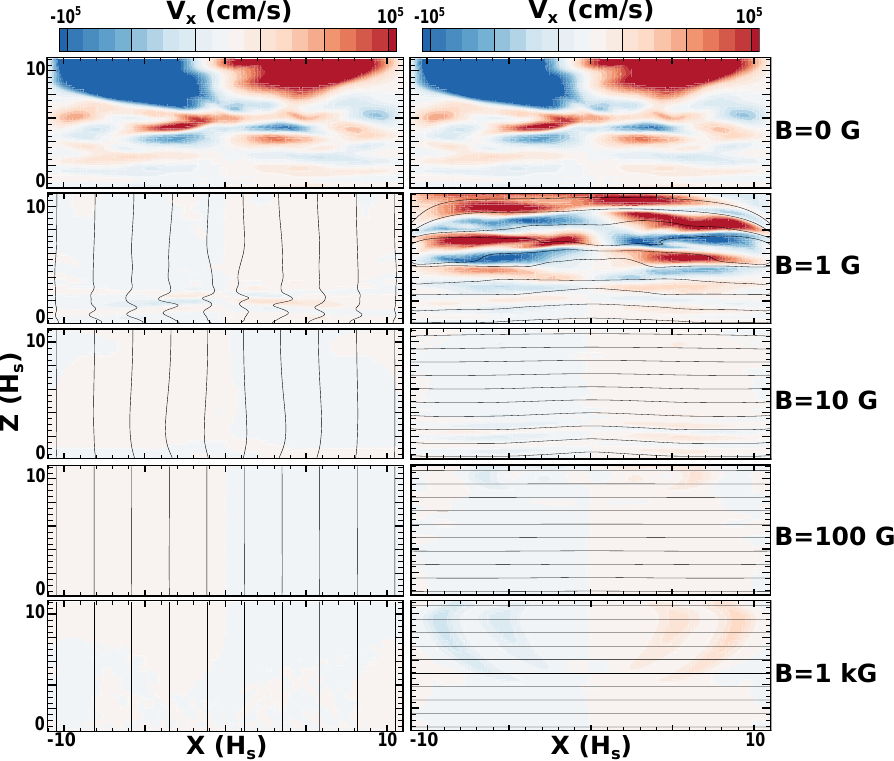}
\caption{Horizontal flow velocity for vertical (left column) and horizontal (right column) magnetic field configurations for all the models shown at the final time, $\mathrm{t}=100~\mathrm{ks}$ except for the non-magnetic case (top row) shown at time $\mathrm{t}=80~\mathrm{ks}$ which is coincidentally more representative than the final time snapshot. The magnetic field strength increases from the top to the bottom.}
\label{fig-vx}
\end{figure*}

To assess the general effects of magnetic fields on atmospheric dynamics, Fig.~\ref{fig-vx} displays horizontal flow velocities across all models at the
conclusion of our simulations. In the left column, we present cases with a
uniform vertical magnetic field orientation, while the right column shows those
with a horizontal magnetic field. The results indicate a clear trend: a strong
vertical magnetic field effectively suppresses horizontal flows by stabilizing
the atmosphere against lateral motion. In our setup, the magnetic field energy
density overcomes gas energy density already for 1~G magnetic field at the top
of the photosphere. Therefore, already the vertical field with strength of $1~\mathrm{G}$
can suppress the horizontal flow. In contrast, models with strong horizontal magnetic fields produce a gentle, steady horizontal flow or `breeze' of approximately $10^4~\mathrm{cm}~\mathrm{s}^{-1}$. This horizontal flow may play an important role in diffusing and eroding surface abundance spots, thereby reducing their lifetime.

\begin{figure*}
\includegraphics[width=\textwidth]{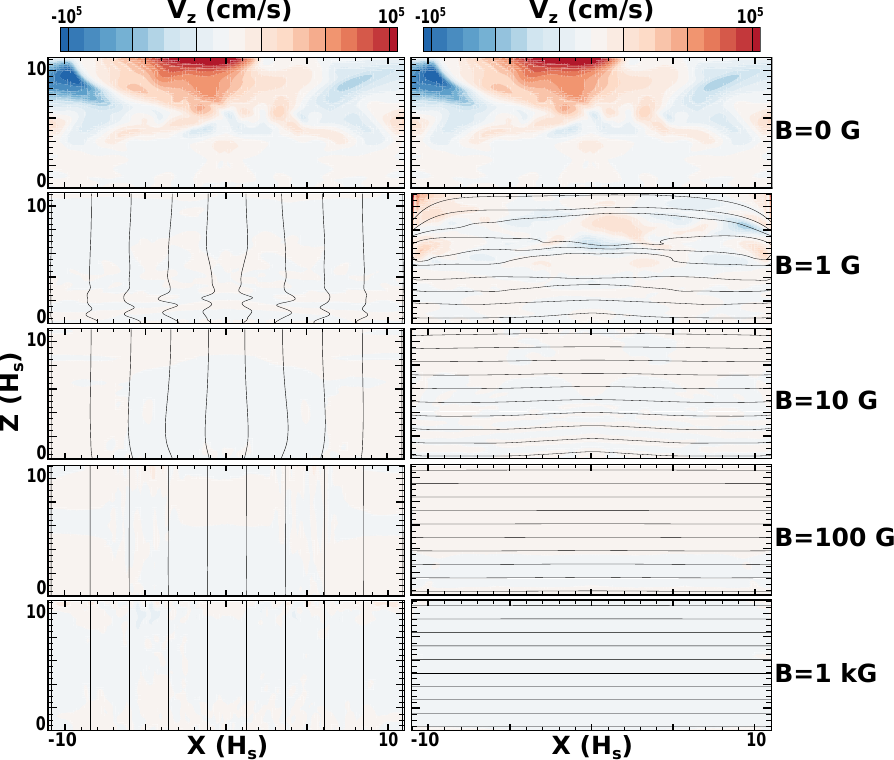}
\caption{The same as figure \ref{fig-vx} but for radial/vertical flow velocity.}
\label{fig-vz}
\end{figure*}

The pattern for vertical flow velocities, shown in Fig.~\ref{fig-vz}, reveals a nearly opposite behavior. Vertical magnetic fields inhibit horizontal flows but allow for limited vertical motion. However, given that the primary temperature gradient in our simulations is oriented horizontally, the absence of sufficient horizontal flow inhibits effective mixing of hot and cool gas, thereby limiting the formation of strong vertical flows.

\begin{figure*}
\includegraphics[width=\textwidth]{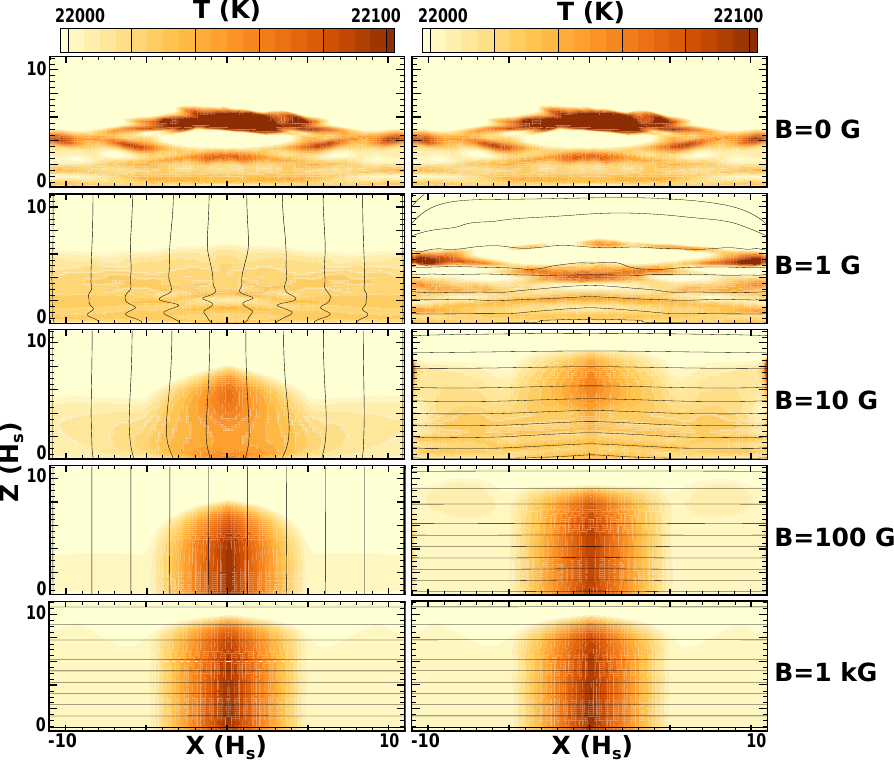}
\caption{The same as figure \ref{fig-vx} but for temperature.}
\label{fig-T}
\end{figure*}

This dynamics is further emphasized in Fig.~\ref{fig-T}, which shows the final temperature distribution for each model. In non-magnetic and weakly magnetized cases (1 and $10~\mathrm{G}$), horizontal flows facilitate the mixing of hot and cool gas, leading to a more homogenized temperature distribution. In contrast, in models with strong magnetic fields, the initial temperature stratification is largely preserved. Here, the magnetic field prevents substantial lateral mixing, maintaining localized regions of hot and cool gas near the initial configuration \citep{peter}.

These findings underscore the influence of magnetic field orientation and strength on atmospheric dynamics. Strong vertical fields stabilize the atmosphere laterally, while strong horizontal fields promote diffusion. This magnetic regulation of flow has implications for the persistence and evolution of surface abundance features, particularly in magnetic chemically peculiar stars where such abundance anomalies are often observed.

\section{Discussion} 
\label{Discussion}

 Our simulations were calculated for a specific set of stellar parameters corresponding to a particular helium-strong chemically peculiar star. However, the phenomenon studied in this paper is not constrained to this specific set of parameters, but is applicable to all stars with an inhomogeneous surface abundance distribution, albeit with differing strength and timescales. The interplay between radiative diffusion, horizontal flows, and other processes influencing abundance evolution suggests that the underlying mechanisms operate across various stellar types, even though the specific characteristics and evolutionary timescales of abundance inhomogeneities may vary significantly.

In non-magnetic stars of spectral type BA, our simulations presented here estimate the speed of atmospheric
flow connected to horizontal temperature inhomogeneities of the order of
$10^5\,\text{cm}\,\text{s}^{-1}$ (cf. Sect.~\ref{sec:Non-magnetic}). Taking into an account a typical radius of
such star, the mixing timescale is of order of tens of days. 
However, it is conceivable that normal BA stars possess surface magnetic fields with typical strengths of the order of 1~G \citep{vegamag,sirmag,blazalmag}. In such a case
the typical flow velocities become order of magnitude lower (cf. Sect.~\ref{sec:magnetic}), leading to a
typical flow timescales of the order of years. This corresponds to a typical
timescale of the evolution of surface abundance spots in HgMn stars
\citep{briquet2010,korphiphe}, thus highlighting a generally good agreement between our model and observations.

 In general, the timescale of the spot evolution is comparable with the radiative diffusion timescale. However, this timescale strongly depends on the location in the atmosphere and on the properties of a particular element such as local relative abundance. Consequently, the radiative diffusion timescale is as short as days in the upper parts of the atmosphere, while it might take centuries in deep photospheric layers \citep{alesido}. For a specific case of mercury in the continuum-forming photospheric regions, the radiative diffusion timescale is comparable to the order of year timescales of horizontal flows. This implies that the photospheric flows and radiative
diffusion operate on similar timescales in HgMn stars, therefore, the horizontal
flow could be one of the processes that precludes strong abundance anomalies in
stars with weak magnetic field.  Although such flows may contribute to dynamical
evolution of abundance spots,  we caution that  timescale to build the peculiar abundances could be longer.

This suggests that while the radiative diffusion timescale may be relatively short in certain regions, the actual process of creating abundance inhomogeneities involves additional factors that extend the overall timescale. These factors include complex interactions between photospheric flows and radiative diffusion, as well as the slower processes of element migration and accumulation. As a result, the timescale for the creation of abundance inhomogeneities might be longer than the radiative diffusion timescale alone would suggest. Understanding these extended timescales is crucial for accurately modeling and interpreting the abundance patterns observed in HgMn stars.

For mass-loss rates predicted in a magnetic helium-strong star HD~37776 \citep{metuje}, the wind velocity in the continuum forming regions of the photosphere, which is of the order of $10\,\text{cm}\,\text{s}^{-1}$, is lower than the velocity of horizontal flows. However, the stellar wind quickly accelerates and overcomes the horizontal flow velocity by one to two orders of magnitude in the upper parts of the photosphere.

On the other hand, our simulations show that magnetic fields stronger than about 100\,G inhibit any flow
generated by the horizontal temperature gradients. These are typical fields
detected in magnetic chemically peculiar stars \citep{morbob,wamimes,grunmimes}.
Therefore, inhibition of horizontal flow by vertical magnetic field contributes
to the establishment of abundance anomalies and their stability. Still, even in
these stars relative rapid flows with speeds of the order of
$10^3\,\text{cm}\,\text{s}^{-1}$ are possible in the regions with horizontal
magnetic fields. Such flow can contribute to the stratification of chemical
elements in magnetic chemically peculiar stars.

The atmospheric flows induced by horizontal temperature gradients may reach up
to $10^5\,\text{cm}\,\text{s}^{-1}$. Such velocities are in principle at the
verge of detection using the method of Doppler imaging. Future observational campaigns might help us verify the predictions of our simulations.

\section{Summary and future work}
\label{summary}
Our study investigates the effects of magnetic fields on atmospheric dynamics in
chemically peculiar stars, focusing on how different magnetic field strengths and orientations
influence horizontal and vertical flows. The photospheric flows are supposed to
originate due to horizontal temperature inhomogeneities connected with abundance
stratification. We find that strong vertical magnetic fields inhibit horizontal flows, preserving temperature stratification and stabilizing abundance anomalies. Conversely, strong horizontal magnetic fields enable a lateral `breeze,' which can erode surface abundance spots over time. In weak or non-magnetic cases, horizontal flow effectively mixes hot and cool gas, resulting in a more homogenized temperature distribution.

Our results for the flow velocity around $10^4~\mathrm{cm}~\mathrm{s}^{-1}$, leading to flux timescales of the order of a year, are in good agreement with the typical timescale for the evolution of surface abundance spots in HgMn stars obtained from observations. Our simulations also confirm that the week horizontal flow can contribute to the dynamical evolution of abundance spots in HgMn stars. On the other hand, the strong vertical magnetic field ($\sim100~\mathrm{G}$) can inhibit the horizontal flow and initiate the formation and stabilization of abundance anomalies in chemically peculiar stars.

Future work should expand these simulations to full 3D to capture the complete
complexity of flow dynamics and magnetic interactions in stellar atmospheres of chemically peculiar stars.
More realistic simulations also require a more detailed treatment of energy
equation accounting for the radiative equilibrium. Additionally, relaxing the planar atmosphere assumption and incorporating a more realistic, curved stellar surface will allow us to more accurately model atmospheric flows and temperature distributions. These improvements will provide deeper insights into the evolution of surface abundance features in magnetic chemically peculiar stars.

\begin{acknowledgements}
A.u.-D. acknowledges NASA ATP grant number 80NSSC22K0628 and support by NASA through Chandra Award number TM4-25001A issued by the Chandra X-ray Observatory 27 Center, which is operated by the Smithsonian Astrophysical Observatory for and on behalf of NASA under contract NAS8-03060. The Astronomical Institute of the Czech Academy of Sciences in Ond\v rejov is supported by the project RVO:67985815.  B.K. acknowledges the support from the Grant Agency of the Czech Republic (GA\^CR 22-34467S).

\end{acknowledgements}

\bibliographystyle{aa}
\bibliography{papers}

\end{document}